\documentclass{article}
\usepackage{amsmath}
\usepackage{bm}
\usepackage{authblk}
\usepackage{graphicx}
\usepackage{xcolor}
\usepackage{url}
\usepackage[inline]{enumitem}
\usepackage{latexsym}
\usepackage{mathptmx}
\usepackage{amsfonts}
\usepackage{amssymb}
\usepackage{mathptmx}
\usepackage{paralist}
\usepackage{amsthm}
\usepackage{comment}
\usepackage{hyphenat}
\usepackage{csquotes}
\usepackage{paralist}
\usepackage[pdftex,colorlinks=true,urlcolor=blue,citecolor=black,anchorcolor=black,linkcolor=black,bookmarks=true]{hyperref}

\usepackage{cleveref}

\usepackage[ruled,vlined,linesnumbered]{algorithm2e}

\newtheoremstyle{scsthe}
{8pt}
{8pt}
{\it}
{}
{\bf}
{.}
{.5em}
{}

\definecolor{mypink}{rgb}{0.858, 0.188, 0.478}

\theoremstyle{scsthe}

\DeclareMathOperator*{\argmax}{arg\,max\;}

\newcommand{\footurl}[1]{\footnote{\url{#1}}}

{}
\newtheorem{remark}{Remark}{}
\newtheorem{example}{Example}{}

\crefname{assumption}{Assumption}{Assumptions}

\crefname{algocf}{algorithm}{algorithms}
\Crefname{algocf}{Algorithm}{Algorithms}

\begin{document}

\title{Model-Based Monitoring and State Estimation for Digital Twins: The Kalman Filter}


\author[$\dagger$]{Hao Feng}
\author[$\dagger$]{Cláudio Gomes}
\author[$\dagger$]{Peter Gorm Larsen}
\affil[$\dagger$]{DIGIT, Department of Electrical and Computer Engineering, Aarhus University, Denmark}

\maketitle
\section*{Abstract}

A digital twin (DT) monitors states of the physical twin (PT) counterpart and provides a number of benefits such as advanced visualizations, fault detection capabilities, and reduced maintenance cost. It is the ability to be able to detect the states inside the DT that enable such benefits. In order to estimate the desired states of a PT, we propose the use of a Kalman Filter (KF). In this tutorial, we provide an introduction and detailed derivation of the KF. We demonstrate the use of KF to monitor and anomaly detection through an incubator system. Our experimental result shows that KF successfully can detect the anomaly during monitoring.

\textbf{Keywords:} state estimate, kalman filter, runtime monitoring, anomaly detection

\section{Introduction}
Since systems are getting increasingly complex, models are needed to better comprehend their behaviors. 
Models can assist with understanding and optimizing the overall system, discovering causes and effects, measuring consequences of changes, and communicating among engineers. 
Once the system is deployed, the models used to build it should not be discarded. 
Instead, they should be integrated into a DT. 

A DT is a system that monitors its PT counterpart, and can reconfigure it \cite{Wright2020,Tao2019a}.
The DT employs calibration algorithms to update the models of the PT with estimates of the parameters.
The calibration is carried out through (co-)simulation inside the DT.
The resulting data is then used to inform experts that check whether the system is performing safely and optimally.
DTs enable advanced visualizations, reconfigurability, safety, predictability, and reduced maintenance. 
However, these benefits do not come without challenges. 

Accurately estimating the states of a PT is the basis for any other visualization and self-adaptation algorithms, but it needs to overcome noise in data, network delays, and the potential lack of sensor signals. 
Fortunately, the use of prediction models enables us to overcome these challenges. This can be achieved by correlating the measured behaviors of the PT with the one predicted by the model, as well as by estimating missing measurements based on the existing ones. 
To this end, the Kalman Filter (KF) \cite{thrun2005probabilistic} can be used. 
The KF provides a way to estimate the state of a dynamic system based on its past behavior, as it is expressed by multiple sequential measurements from sensors, that is better than the estimate obtained by using only one measurement alone.

In this tutorial, we provide an introduction and detailed derivation of the KF, assuming that the reader is familiar with basic statistics.
We use, as a running example, an incubator system \cite{feng2021}: a simple styrofoam insulated container with the ability to reach a certain temperature within a box and regulate it regardless of the content. We demonstrate the use of KF to estimate the states of the air inside the incubator and the heatbed (a heating unit inside the incubator) in the scenario that only the measurement data of the air inside the box is accessible.

The ability to predict the states of the system also enables anomaly detection. This can be done by the DT by detecting that the predictions from its models are no longer aligned with the data from the PT indicate.
In our work, we opened the lid of the incubator, which we can detect as a fault in the system. 
By comparing the states estimated by the KF and the measurements from the sensor, the fault is detected.

The rest of the paper is organized as follows: \cref{sec:incubator_system} describes the incubator system including the PT and the DT, and \cref{sec:kf} presents the derivation of the KF. \Cref{sec:anomaly_detection} demonstrates the use of KF for anomaly detection in the incubator system. Finally, we provide the conclusions in \cref{sec:conclusion}.

\section{Incubator Digital Twins} \label{sec:incubator_system}

Incubator DTs system consists of a PT and a corresponding DT. 
The DT replicates the essential features of the PT and enables some techniques increasing the value of the PT.

\begin{figure}[h!]
	\centering
	\includegraphics[scale=0.85]{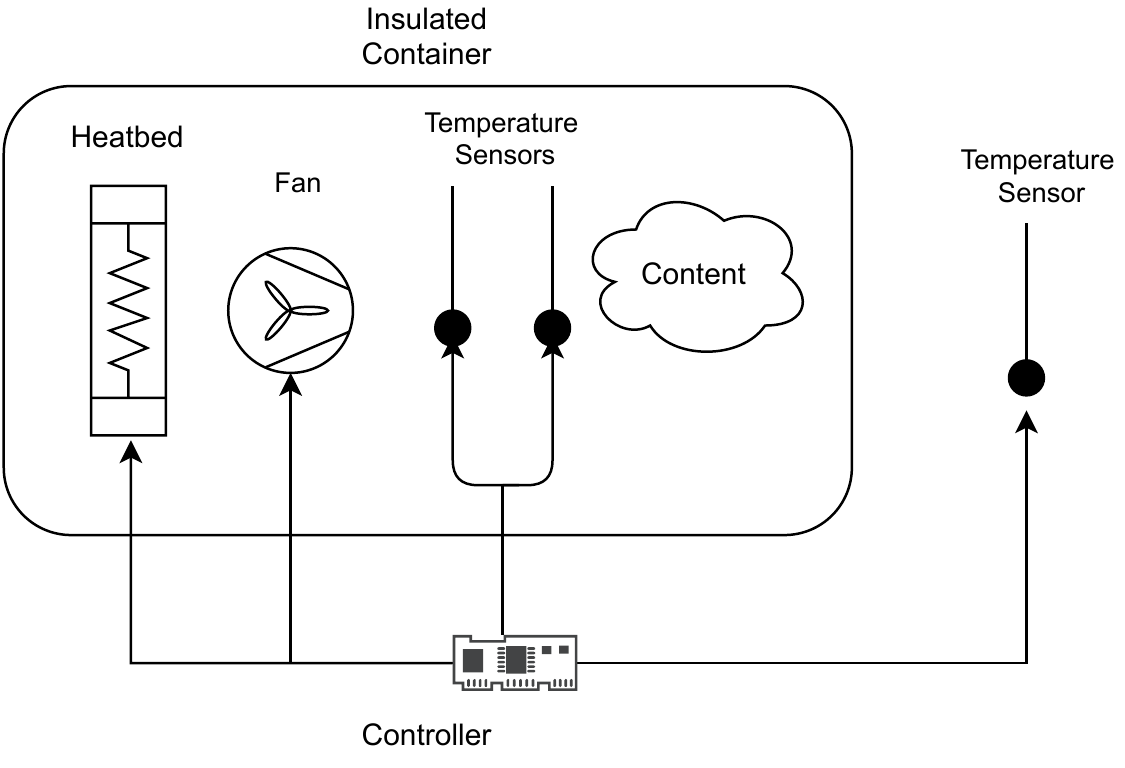}
	\caption{Schematic overview of the Incubator.}
	\label{fig:incubator}
\end{figure}

\paragraph{Physical Twin.}
The incubator is a system with the ability to hold a desired temperature within an insulated container.
The physical components include a controller and a plant. The controller is running on a Raspberry PI, and the plant includes a Styrofoam box, a heater, a fan, and three temperature sensors.
The overall systematic diagram of the Incubator is shown in \cref{fig:incubator}.
For more details, please refer to the report in \cite{feng2021} and the online documentation at \url{https://github.com/INTO-CPS-Association/example-incubator}.

\paragraph{Digital Twin}
In a DT, we highlight the following important parts \cite{Wright2020} (inspired by autonomic computing \cite{Kephart2003}): 
\begin{inparadesc} 
	\item[Data:] collected from the PT through sensors and actuators over time;
	\item[Models:] representing knowledge about different aspects of both the cyber and the physical of the PT and its environment; and
	\item[Algorithms:] representing techniques that use data and models, manipulating those to generate more data and knowledge (e.g., fault detectors, supervisory controllers, state estimators, optimizers).
\end{inparadesc} 

The data in the PT are transmitted through the Raspberry PI using a RabbitMQ server, which makes it easily accessible to other algorithms. 
In order to study the dynamics of temperature inside the incubator, a dynamic model is necessary. It is worth noting that models can be described at different levels of abstraction, and typically, accurate models are also very slow to simulate.
We experimented with different models, which can be found online, and selected the one that offered the best compromise between accuracy and performance.

Usually a physical dynamic system is modeled in the form of differential equations organized in a state-space model in continuous time.
However, in order to represent such system as a discrete process that produces samples at some finite frequency, it is necessary to transform the continuous-time state space model to a discrete-time state space model, based on a well known process called discretization.
We therefore focus on linear discrete-time dynamical systems, and refer the reader to \cite{Li2005} for details on discretization.

The general form of a linear discrete-time dynamical system is
\begin{equation} \label{equ:state_space_equation_discrete}
		{x}_k = A x_{k-1} + B u_k, \hspace{3em}
		y_k =C x_k,
\end{equation}
where $x_k$ and $x_{k-1}$ are state vectors, $u_k$ is the input vector, and $y_k$ is the measurement vector. 
The subscripts indicate the time step. 
In addition, $A$, $B$, and $C$ are matrices expressing the rule of how the system states advances and how the measurements relate to the state. 

The dynamics model described below mainly focuses on describing the temperature behaviors inside the Styrofoam box, and assume that:
the temperature inside the box is distributed uniformly; 
the box walls do not accumulate heat; 
and the specific heat capacity of the air inside the box is constant.
For the full definition of the model please refer to \cite{feng2021}.

\begin{example}\label{ex:incubator}
  The incubator system is described by the following linear discrete time system:
  \begin{equation} \label{equ:model_b}
     	\begin{bmatrix}
     		T^{h}_k\\
     		T^{b}_k
     	\end{bmatrix}
     	=
     	A
     	\begin{bmatrix}
     		T^{h}_{k-1}\\
     		T^{b}_{k-1}
     	\end{bmatrix}
     	+ B
     	\begin{bmatrix}
     		P_k\\
     		T^r_k
     	\end{bmatrix}, \hspace{3em}
    y_k =	
    	\begin{bmatrix}
    		0 & 1
    	\end{bmatrix}
    		\begin{bmatrix}
    				T^{h}_k\\
    				T^{b}_k
    		\end{bmatrix}
        = T^{b}_k,
  \end{equation}
  where $T^{h}$, $T^{b}$, and $T^{r}$ are the temperature of heater, the temperature of the air inside the incubator, and the temperature of the room, respectively. $P$ represents the power supply on/off function. $A$ and $B$ are $2$ by $2$ matrices. 
\end{example}

As can be seen from \cref{ex:incubator}, the measurements consist only of the temperature $T^{b}$, but the state of the model consists of both $T^{b}$ and $T^{h}$. 
We will now explore different ways to estimate the state in \cref{ex:incubator}.

\paragraph{State Estimation under Exact State.}
Let us assume that we know exactly the initial state of the system $T^{b}_0,T^{h}_0$, and the initial input $P_1, T^r_1$.
We can then simply apply the expressions in \cref{ex:incubator} to obtain the next predicted state $T^{b}_1,T^{h}_1$.
If we want to know the $k$-th state, and we have the history of inputs, denoted as $u_{1:k}=u_1,u_2,u_3,\ldots,u_{k-1}$, then we simply have to apply the expressions in \cref{ex:incubator} iteratively, from the given initial state.
This is no different that simulating the system.

\paragraph{State Estimation under Uncertain State with No Measurement.}
Now imagine that the initial state $T^{b}_0$ is known exactly, but $T^{h}_0$ is not known exactly. 
Instead, we know that it follows a Gaussian distribution with mean $\mu_{T^{h}_0}$, and variance $\Sigma_{T^{h}_0}$:
$
T^{h}_0 \sim \mathcal{N}(\mu_{T^{h}_0},\Sigma_{T^{h}_0}).
$
Suppose we want to estimate the most likely state $T^{b}_1,T^{h}_1$.
If we have no access to output measurements, then we could simply apply the expressions in \cref{ex:incubator} to the most likely initial state.
Since we know that the most likely output of a Gaussian process is its mean (recall that the Gaussian distribution resembles a bell-shaped distribution), we would then apply \cref{equ:model_b} to the initial state $\mu_{T^{h}_0},T^{b}_0$.

\paragraph{State Estimation under Uncertain State with Measurement.}
Following on the previous paragraph, imagine that a measurement for $T^{b}_1$ is given, in addition to the initial state $T^{b}_0$ and inputs $P_1, T^r_1$. Can we do better than just use $\mu_{T^{h}_0}$?
The answer is yes, because $T^{b}_1$ provides us with some extra information to try to find what the real initial state was.
We can insert the known values into \cref{equ:model_b} and estimate the unknown value $T^{h}_0$.
Expanding the $2 \times 2$ matrices, and focusing on the expression $T^{b}_1$, we obtain
$
T^{b}_1 = A_{21} T^{h}_0 + A_{22} T^{b}_0 + B_{21} P_1 + B_{22} T^r_1,
$
which can be solved to yield the solution to $T^{h}_0$. Then we insert the values of $T^{h}_0$, $T^{b}_0$, and inputs $P_1, T^r_1$ into \cref{equ:model_b} to acquire the value of $T^{h}_1$. 
Here we can see that $T^{h}_1$ is dependent on $T^{b}_1$. 
If the measurement $T^{b}_1$ is noisy, then $T^{h}_1$ is noisy as well. 
In addition, if the model contains process noises as well, this makes the uncertainty of $T^{h}_1$ larger. 
In practice, it is not likely for measurements and models to be noise free.
This leads to the third situation where measurements and models are noisy.

\paragraph{State Estimation under Process Noise.}
Now imagine that there is noise in the state transition function, and in the measurements, as summarized in the following example.

\begin{example}\label{ex:incubator_noise}
  The incubator system with noise is described by the following linear discrete time system:
  \begin{equation} \label{equ:model_b_noisy}
     	\begin{bmatrix}
     		T^{h}_k\\
     		T^{b}_k
     	\end{bmatrix}
     	=
     	A
     	\begin{bmatrix}
     		T^{h}_{k-1}\\
     		T^{b}_{k-1}
     	\end{bmatrix}
     	+ B
     	\begin{bmatrix}
     		P_k\\
     		T^r_k
     	\end{bmatrix} 
      + \epsilon_k, \hspace{3em}
    y_k = T^{b}_k + \delta_k,
  \end{equation}
  where $\epsilon_k$ and $\delta_k$ are random variables representing process noise and measurement noise, satisfying $\epsilon_k \sim \mathcal{N}(0,R_k)$ and $\delta_k \sim \mathcal{N}(0,Q_k)$, respectively.
\end{example}

Given values for the initial states $T^{b}_0,T^{h}_0$, inputs $P_1, T^r_1$, and measurement $T^{b}_1$, we want to estimate the value of $T^{h}_1$.
In this case we cannot simply apply \cref{equ:model_b_noisy} because 
the measurement and the model are noisy. 

\paragraph{Goal of the Kalman Filter.}
Applied to the incubator case, the KF combines measurements $T^{b}_k$,
inputs $P_k, T^r_k$, 
Gaussian distribution parameters $\mu_{T^{h}_{k-1}},\Sigma_{T^{h}_{k-1}}$,
and a model in \cref{equ:model_b_noisy}, to estimate the parameters 
$\mu_{T^{h}_{k}},\Sigma_{T^{h}_{k}}$:
\begin{equation} \label{equ:kf_summary}
  \begin{bmatrix}
		\mu_{T^{h}_{k}}\\
		\Sigma_{T^{h}_{k}}
	\end{bmatrix}
	=
	\mathit{KF}(
  	\begin{bmatrix}
  		\mu_{T^{h}_{k-1}}\\
  		\Sigma_{T^{h}_{k-1}}
  	\end{bmatrix},
  	\begin{bmatrix}
  		P_k\\
  		T^r_k
  	\end{bmatrix},
    T^{b}_k
  )
  \text{, \ \ \ with }
  T^{h}_K \sim \mathcal{N}(\mu_{T^{h}_k},\Sigma_{T^{h}_k}).
\end{equation}
In the following section, we discuss how to derive KF. 

\section{Kalman Filter for state estimation} \label{sec:kf}

We derive the KF from the perspective of probability theory.
We therefore start with a small review of the background concepts. 
The reader who is familiar with probability theory may skip them and resume in \cref{sec:linear_system}.

\subsection{Basic Probability Theory}
\label{sec:background_probability}

A random variable, for example $X$, may be assigned a value representing the outcomes of an experiment. 
We use $p(X=x)$ to denote the probability that the random variable $X$ has value $x$. 
The probabilities of all outcomes sum up to $1$. 
Sometimes we write $p(x)$ instead of $p(X=x)$, for brevity.

As we will show later, one of the important problems in state estimation is the computation of a conditional probability function. 
For instance, if $x$ is the state vector we want to estimate, and $y$ is the measured data, the conditional probability $p(X=x|Y=y)$ (or $p(x|y)$) represents what is the probability of state $X=x$ given the measurement $y$.
The following represent well known equations involving the conditional probability. 
\begin{equation}
	\label{equ:conditional_prob}
	p(x|y) = \frac{p(x\cap y)}{p(y)} = \frac{p(x, y)}{p(y)}, \hspace{3em}
  p(x|y) = \frac{p(y|x)p(x)}{p(y)}.     
\end{equation}
where 
$p(x)$ is called the prior (probability of state $X=x$ without any additional information);
$p(x|y)$ is called the posterior;
$p(y|x)$ is called the likelihood; and
$p(y)$ is called the marginal likelihood which is the total probability of observing $Y=y$.

\begin{remark}\label{remark:ct_prob}
	For a continuous random variable, we use $p(x)$ to denote a Probability Density Function (PDF) of random variable $X=x$ and the probability of $X=x$ is $Pr(x)=p(x)\Delta x$, where $\Delta x > 0$ is a sufficient small value. 
	In this tutorial, we will mostly use $p(x)$ to represent the PDF for a continuous random variable. 
\end{remark}

The joint probability can be generalized to multiple random variables. 
According to the definition of conditional probability in \cref{equ:conditional_prob}, we obtain
\begin{subequations} \label{equ:joint_prob_multiple_random_variables}
	\begin{align}
		p(x_n|x_{n-1},\ldots,x_1) &= p(x_n,x_{n-1},\ldots,x_1)/p(x_{n-1},\ldots,x_1)\\
		p(x_n,x_{n-1},\ldots,x_1)&=	p(x_n|x_{n-1},\ldots,x_1)p(x_{n-1},\ldots,x_1).\label{equ:joint_prob_expand_rule}
	\end{align}
\end{subequations}

An example of a PDF for a continuous random variable is the Gaussian distribution, commonly used to model process and measurement noise.
One of the advantages of Gaussian distribution is that if the prior and the likelihood are Gaussian distributed, the posterior is also Gaussian distributed. Such distribution makes the calculation of expectation and variance simpler, which makes possible the derivation of the KF.
The univariate Gaussian distribution with a mean $\mu$ and variance $\sigma^2$ is defined as
\begin{equation}\label{equ:gaussian}
	p(x)=(2\pi \sigma^2 )^{-\frac{1}{2}}
	\exp\{-\frac{1}{2} \frac{(x-\mu)^2}{\sigma^2}\}.
\end{equation}
If the mean $\mu$ and variance $\sigma^2$ are unknown, we define a conditional probability of $X=x$ conditioned on $\mu$ and $\sigma^2$ with the notation $p(x|\mu,\sigma^2)$.
In addition, we write
$X \sim \mathcal{N}(\mu,\sigma^2)$
to denote that $X$ follows a Gaussian distribution with a mean $\mu$ and variance $\sigma^2$.
The multivariate Gaussian distribution is characterized by a mean vector $\mu$ with the same dimension of state vector $x$ and a covariance matrix $\Sigma$, having the form of  
\begin{equation} \label{equ:multivariate_gaussian}
	p(x)=\det (2\pi \Sigma)^{-\frac{1}{2}}
	\exp\{-\frac{1}{2}(x-\mu)^T\Sigma^{-1}(x-\mu)\}.
\end{equation}

\subsection{Linear Discrete-Time Dynamical System with Gaussian Uncertainty} \label{sec:linear_system}

As introduced in \cref{sec:incubator_system}, and exemplified in \cref{ex:incubator_noise}, the general form of a linear discrete-time dynamical system with noise is given as:
\begin{equation}\label{equ:state_space_equation_noise}
		{x}_k =Ax_{k-1}+Bu_k+\epsilon_k, \hspace{3em} 
    y_k =Cx_k+\delta_k,
\end{equation}
where $x_k$ and $x_{k-1}$ are state vectors, $u_k$ is the input vector, and $y_k$ is the measurement. 
This means that previous states such as $x_{k-2}$ or $x_{k-8}$ do not contribute directly to the calculation of state $x_k$ and measurement $y_k$, but they contributed implicitly through $x_{k-1}$. 
In addition, we have the following assumption:
\begin{compactitem}
\item \Cref{equ:state_space_equation_noise} is the best predictor of state $x_k$.
\item The process error $\epsilon_k$ and measurement error $\delta_k$ satisfy $\epsilon_k \sim \mathcal{N}(0,R_k)$ and $\delta_k \sim \mathcal{N}(0,Q_k)$, respectively.
\item The initial state $x_0$ satisfies $x_0 \sim \mathcal{N}(\mu_{x_0},\Sigma_{x_0})$.
\end{compactitem}

We now proceed to show that, under the above assumptions, all quantities in \cref{equ:state_space_equation_noise} follow Gaussian distributions.
This is an important result because it is very easy to obtain the most likely value of a Gaussian distribution, i.e. its mean.
If presented with an analytical expression as in \cref{equ:gaussian}, we can obtain the mean by differentiating the expression and equating it to zero.
Moreover, the covariance is obtained by calculating the second derivative of \cref{equ:gaussian}.

We rewrite \cref{equ:state_space_equation_noise} in terms of $\epsilon_k$ and $\delta_k$:
\begin{equation} \label{equ:transformation_noise}
		\epsilon_k  = {x}_k - (Ax_{k-1}+Bu_k), \hspace{3em}
		\delta_k = y_k - C x_k.
\end{equation}
According to \cref{equ:multivariate_gaussian}, the PDF of $\delta_k$ is
\begin{equation} \label{equ:noise_distribution}
	p(\delta_k) = \det (2\pi Q_k)^{-\frac{1}{2}}\exp\{ -\frac{1}{2}(\delta_k-0)^TQ_k^{-1}(\delta_k-0)\}.
\end{equation}
Substituting \cref{equ:transformation_noise} into \cref{equ:noise_distribution} yields
\begin{equation}
	p(\delta_k) = p(y_k - C x_k) 
	= \det (2\pi Q_k)^{-\frac{1}{2}}\exp\{ -\frac{1}{2}(y_k -Cx_k)^TQ_k^{-1}(y_k -Cx_k)\}
\end{equation}
Which is the expression for the Gaussian distribution $y_k \sim \mathcal{N}(Cx_k,Q_k)$.
Since $x_k$ is an unknown, we write $p(y_k | x_k)$.
Applying the same reasoning to $\epsilon_k$, we get the expression for $x_k \sim \mathcal{N}(Ax_{k-1}+Bu_k,R_k)$.
This shows that the states, measurements, and noise, are all Gaussian distributions.

\begin{remark}\label{rem:distributions}
	All variables in \cref{equ:state_space_equation_noise} are Gaussian distributed, with parameters:
	\begin{equation}
		\begin{split}
			\epsilon_k \sim \mathcal{N}(0,R_k)\\
			\delta_k \sim \mathcal{N}(0,Q_k)\\
			x_k \sim \mathcal{N}(Ax_{k-1}+Bu_k,R_k)\\
			y_k \sim \mathcal{N}(Cx_k,Q_k)
		\end{split}
	\end{equation}
\end{remark}

\subsection{Maximum A Posteriori Estimation}
\label{sec:map}

One of the tasks of state estimation is to calculate the most likely value of an unknown quantity, given its probability distribution (recall \cref{sec:incubator_system}).
For example, in the incubator system, the unknown quantity can be the state of the incubator system (temperature of the heater and the air inside the box), and the known quantity is the measured air temperature.
Maximum a posteriori (MAP) estimate is the estimation of an unknown quantity from a distribution of the unknown quantity. 
It is the most probable state $x$ given measurement data $y$, formulated as:
\begin{equation} \label{equ:map_pre}
	\hat{x} = \argmax_{x} p(x|y).
\end{equation}
Note that, according to \cref{remark:ct_prob}, when the random variable $X$ is continuous, the estimation becomes $\hat{x} = \argmax_{x} p(x|y)\Delta x$.
However, since $\Delta x$ does not depend on $x$, $\argmax_{x} p(x|y)\Delta x$ is equivalent to $\argmax_{x} p(x|y)$.

If we do not have an analytical expression for \cref{equ:map_pre}, we may attempt to reformulate it into an equivalent problem, comprised of terms for which we have analytical expressions.
Substituting Bayes rule in \cref{equ:conditional_prob} into \cref{equ:map_pre}, we obtain the MAP objective function
\begin{equation} \label{equ:map_middle}
	\hat{x} = \argmax_{x} \frac{p(y|x)p(x)}{p(y)},
\end{equation}
and because $p(y)$ is independent of $X$, \cref{equ:map_middle} is simplified to
\begin{equation} \label{equ:map}
	\hat{x} = \argmax_{x} p(y|x) p(x).
\end{equation} 

\subsection{Objective Function of Kalman Filter}

Suppose we have collected a series of historical inputs, $u_{1:k}=u_1,u_2,u_3,\ldots,u_{k-1}$, and measurement outputs, $y_{1:k-1}=y_1,y_2,y_3,\ldots,y_{k-1}$, from a linear discrete time system as in \cref{equ:state_space_equation_noise}.
At time step $k$, we apply an input $u_k$ to the system. 
Then the state of the system advances to the state $x_k$ from state $x_{k-1}$, and we obtain measurement data $y_k$. 
The ultimate outcome of the KF is the most probable state $x$ using the posterior distribution given all data.

Applying the MAP discussed in \cref{sec:map}, we formulate the following
\begin{equation} \label{equ:bel_xt}
	\hat{x}_k = \argmax_{x_k} p(x_k|y_{1:k},u_{1:k}).
\end{equation}

In what follows, we describe the procedure to turn \cref{equ:bel_xt} into a form similar to \cref{equ:map}.
Based on \cref{equ:joint_prob_multiple_random_variables} and \cref{equ:map}, we expand \cref{equ:bel_xt} to
\begin{equation} \label{equ:xt_beyasian_inference}
	\begin{split} 
		\hat{x_k} &= \argmax_{x_k} p(x_k|y_k, y_{1:k-1},u_{1:k}) 
		 =\argmax_{x_k} \frac{p(y_k|x_k,y_{1:k-1},u_{1:k})p(x_k|y_{1:k-1},u_{1:k})}{p(y_k|y_{1:k-1},u_{1:k})}\\
		&=\argmax_{x_k} p(y_k|x_k,y_{1:k-1},u_{1:k})p(x_k|y_{1:k-1},u_{1:k}).
	\end{split}
\end{equation}

According to \cref{rem:distributions}, the distribution of current measurement $y_k$ only depends on the current state $x_k$.
We can therefore conclude that
$p(y_k|x_k,y_{1:k-1},u_{1:k}) = p(y_k|x_k)$
and simplify \cref{equ:xt_beyasian_inference} to:
\begin{equation} \label{equ:map_states}
	\hat{x_k} = \argmax_{x_k} p(y_k|x_k)p(x_k|y_{1:k-1},u_{1:k}).
\end{equation}

Our goal is to derive a recursive calculation of the KF.
So we aim at reducing the problem in \cref{equ:bel_xt} to something that involves the term $p(x_{k-1}|y_{1:k-1},u_{1:k-1})$.
In \cref{equ:map_states}, the term $p(y_k|x_k)$ is already at its simplest form, since it matches exactly the distributions in \cref{rem:distributions}, so we will focus on the term $p(x_k|y_{1:k-1},u_{1:k})$.
Based on \cref{rem:distributions}, we can see that $x_k$ depends on $x_{k-1}$ and $u_k$, so we are looking for a way to relate $y_{1:k-1}$ to $x_{k-1}$.
This is where Chapman-Kolmogorov equation \cite{ross2014a} comes into play.
Applying this equation to $p(x_k|y_{1:k-1},u_{1:k})$ gives:
\begin{equation}\label{eq:chapman}
		p(x_k|y_{1:k-1},u_{1:k}) = \int p(x_k, x_{k-1}|y_{1:k-1},u_{1:k})\,dx_{k-1}.
\end{equation}
Applying Bayes theorem, the term inside the integral can be simplified to 
\begin{equation}
	\begin{split}
		p(x_k, x_{k-1}|y_{1:k-1},u_{1:k}) &= p(x_k|x_{k-1},y_{1:k-1},u_{1:k})p(x_{k-1}|y_{1:k-1},u_{1:k})\\ & =
    p(x_k|x_{k-1},u_{k})p(x_{k-1}|y_{1:k-1},u_{1:k}).
	\end{split}
\end{equation}
Because future input $u_k$ can not affect the past state $x_{k-1}$, we can safely omit it from $p(x_{k-1}|y_{1:k-1},u_{1:k})$, therefore obtaining the term $p(x_{k-1}|y_{1:k-1},u_{1:k-1})$.
To summarise, \cref{eq:chapman} can be simplified to:
\begin{equation} \label{eq:chapman_2}
	p(x_k|y_{1:k-1},u_{1:k}) = \int p(x_k|x_{k-1},u_{k})p(x_{k-1}|y_{1:k-1},u_{1:k-1}) \,dx_{k-1}.
\end{equation}

Finally, we insert \cref{eq:chapman_2} into \cref{equ:map_states}, obtaining the following equivalent formulation:
\begin{equation} \label{equ:obj_expand}
	\begin{split}
		\hat{x_k} &= \argmax_{x_k} p(x_k|y_{1:k},u_{1:k}) \\ &= \argmax_{x} p(y_k|x_k)\int p(x_k|x_{k-1},u_{k})p(x_{k-1}|y_{1:k-1},u_{1:k-1}) \,dx_{k-1}.
	\end{split}
\end{equation} 

The derivation above shows the objective function for KF based on MAP.
Here we summarize what we know about this problem:
\begin{inparaitem}
\item $p(y_k|x_k)$ is the PDF of $\mathcal{N}(Cx_k,Q_k)$ (\cref{rem:distributions}); and
\item $p(x_k|x_{k-1},u_{k})$ is the PDF of $\mathcal{N}(Ax_{k-1}+Bu_k,R_k)$ (\cref{rem:distributions}).
\end{inparaitem}
In addition, here is what we need to know in order to solve the problem: does $p(x_k|y_{1:k},u_{1:k})$ match the PDF of a Gaussian distribution? If it does, the solution is just its mean.

In the interest of solving the problem, let us for now assume that $p(x_k|y_{1:k},u_{1:k})$ indeed matches the PDF of a Gaussian distribution with mean $\mu_{x_{k}}$ and $\Sigma_{x_{k}}$:
$\mathcal{N}(\mu_{x_{k}}, \Sigma_{x_{k}})$.
It follows that the term $p(x_{k-1}|y_{1:k-1},u_{1:k-1})$ in \cref{equ:obj_expand} matches the PDF of $\mathcal{N}(\mu_{x_{k-1}}, \Sigma_{x_{k-1}})$.
As \cref{sec:prediction_phase} will show, these facts are instrumental in showing that the integral term in \cref{equ:obj_expand} will match the PDF of a Gaussian distribution scaled by a factor $\eta$, with mean and covariance as follows: $ \mathcal{N}( \overline{\mu}_{x_k}=Bu_k + A\mu_{x_{k-1}} , \overline{\Sigma}_{x_k} = (R_k + A\Sigma_{x_{k-1}}A^T)^{-1})$.

Then, in \cref{sec:solving_opt_problem}, we show how under the above conclusions, the mean $\mu_{x_{k}}$ and covariance $\Sigma_{x_{k}}$ are computed, from the knowledge of $\mu_{x_{k-1}}, \Sigma_{x_{k-1}}$.
This calculation is the KF.

\subsection{Prediction Phase} \label{sec:prediction_phase} 

In this section, we focus on the term 
$\int p(x_k|x_{k-1},u_{k})p(x_{k-1}|y_{1:k-1},u_{1:k-1}) \,dx_{k-1}$.  
This term represents the prediction phase of KF because it utilizes the previous estimates state $x_{k-1}$ with state transition function to predict state $x_{k}$.
The term $p(x_{k-1}|y_{1:k-1},u_{1:k-1})$ is the posterior of state $x_{k-1}$ given historical data.

Regarding the initial state $x_0$, the posterior is simply $p(x_0)$ since we do not have access to historical data. According to \cref{equ:obj_expand}, 
the posterior of state $x_1$ is 
$
		p(x_1|y_1,u_{1}) \propto p(y_1|x_1)\int p(x_1|x_{0},u_{1})p(x_{0}) \,dx_{0},
$
where $\propto$ means \emph{proportional to}.
Based on our assumptions in \cref{sec:linear_system}, all the terms, $ p(y_1|x_1)$, $ p(x_1|x_{0},u_{1})$, and $p(x_{0})$ , are Gaussian distributed. 
Thus the posterior of state $x_1$ is also Gaussian distributed. 
By induction, all the posterior of states are Gaussian distributed as well as $p(x_{k-1}|y_{1:k-1},u_{1:k-1})$.

Let us assume the mean and covariance of $p(x_{k-1}|y_{1:k-1},u_{1:k-1})$ to be $\mu_{x_{k-1}}$ and $\Sigma_{x_{k-1}}$.
Furthermore, we use the Gaussian distribution in $\int p(x_k|x_{k-1},u_{k})p(x_{k-1}|y_{1:k-1},u_{1:k-1}) \,dx_{k-1}$ and let $\tilde{x_{k}}=\int p(x_k|x_{k-1},u_{k})p(x_{k-1}|y_{1:k-1},u_{1:k-1}) \,dx_{k-1}$, leading to
\begin{equation} \label{equ:tilde_x}
	\begin{split}
		\tilde{x_{k}} &=\int p(x_k|x_{k-1},u_{k})p(x_{k-1}|y_{1:k-1},u_{1:k-1}) \,dx_{k-1}\\
		&= \gamma \int \exp \! \{
		\begin{aligned}[t]
			 -\frac{1}{2}({x}_k - Ax_{k-1}-&Bu_k)^TR_k^{-1}({x}_k - Ax_{k-1}-Bu_k)\\
			 & -\frac{1}{2}(x_{k-1} -\mu_{x_{k-1}})^T \Sigma_{x_{k-1}}^{-1}(x_{k-1} -\mu_{x_{k-1}})  \}\,dx_{k-1},
		\end{aligned}
	\end{split}
\end{equation}
where $\gamma = \det (2\pi R_k)^{-\frac{1}{2}}\det \Sigma_{x_{k-1}}^{-\frac{1}{2}}$. 
Such $\gamma$ plays the role of normalization. 

Let 
$
	L_k = \frac{1}{2}({x}_k - Ax_{k-1}-Bu_k)^TR_t^{-1}({x}_k - Ax_{k-1}-Bu_k) + \frac{1}{2}(x_{k-1} -\mu_{x_{k-1}})^T  \Sigma_{x_{k-1}}^{-1}(x_{k-1} -\mu_{x_{k-1}}).
$
Then, we have
\begin{equation} \label{equ:predicted_x_k}
	\tilde{x_{k}} =\gamma \int \exp\{-L_k\} \,dx_{k-1}, 
\text{where $L_k$ is quadratic in $x_{k-1}$ and $x_{k}$.}
\end{equation}
We can express $L_k$ in terms of $loss(x_{k-1},x_k)$ and $loss(x_k)$, then we obtain
\begin{equation} \label{equ:L_k}
	L_k = loss(x_{k-1},x_k) + loss(x_k),
\end{equation}
where $loss(x_{k-1},x_k)$ incorporates terms relative to $x_{k-1}$ and $x_k$, $loss(x_k)$ only contains items about $x_k$. 
The decomposition of $L_k$ separates the terms $x_k$ and $x_{k-1}$. 
Such decomposition simplifies \cref{equ:predicted_x_k} to 
\begin{equation} \label{equ:bel_xk_predicted}
	\begin{aligned}
		\tilde{x_{k}} &=\gamma \int \exp\{-L_k\} \,dx_{k-1} =\gamma \int \exp\{-loss(x_{k-1},x_k) - loss(x_k)\} \,dx_{k-1}\\
		&=\gamma \exp\{-loss(x_k)\}  \int \exp\{-loss(x_{k-1},x_k) \} \,dx_{k-1}.
	\end{aligned}
\end{equation}

$loss(x_k)$ and $loss(x_{k-1},x_k)$ are both quadratic in $x_k$ and $x_{k-1}$, which have the form of a Gaussian distribution but without normalization. 
Thus we will rearrange $loss(x_k)$ and $loss(x_{k-1},x_k)$ in terms of mean and covariance of a Gaussian distribution. 

First, we analyze $loss(x_{k-1},x_k)$. 
The covariance of $x_{k-1}$ in $loss(x_{k-1},x_k)$ is the inverse of coefficient of $x_{k-1}^2$ in $L_k$. 
By taking the second derivative of $L_k$ with respect to $x_{k-1}$, we acquire the covariance of $x_{k-1}$ in $loss(x_{k-1},x_k)$. 
Setting the first derivative of $L_k$ with respect to $x_{k-1}$ to zero gives the mean value of $x_{k-1}$ in $loss(x_{k-1},x_k)$. Then, $loss(x_{k-1},x_k)$ is expressed as 
\begin{equation}
	\begin{split}
		\mathit{loss}&(x_{k-1},x_k) =  \frac{1}{2}( x_{k-1} - (A^TR_k^{-1}A+\Sigma_{x_{k-1}}^{-1})	[A^TR_k^{-1}(x_k-Bu_k)+\Sigma_{x_{k-1}}^{-1}\mu_{x_{k-1}}] )^T \\ & (A^TR_k^{-1}A+\Sigma_{x_{k-1}}^{-1})^{-1} 
		(x_{k-1} - (A^TR_k^{-1}A+\Sigma_{x_{k-1}}^{-1})	[A^TR_k^{-1}(x_k-Bu_k)+\Sigma_{x_{k-1}}^{-1}\mu_{x_{k-1}}] ).
	\end{split}
\end{equation}

We know that the integral of a Gaussian distribution of \cref{equ:multivariate_gaussian} should equal to 1. 
Furthermore, the covariance of $x_{k-1}$ in $loss(x_{k-1},x_k)$ is $\Psi = A^TR_k^{-1}A+\Sigma_{x_{k-1}}^{-1}$. Thus we conclude that
$
\int \det\{2\pi \Psi \} ^{-\frac{1}{2} }  \exp\{-L_k(x_{k-1},x_k) \} \,dx_{k-1} = 1
$
and therefore
$
\int  \exp\{-L_k(x_{k-1},x_k) \} \,dx_{k-1} = \det\{2\pi \Psi \} ^{\frac{1}{2} }.
$
Because $\det\{2\pi \Psi \} ^{\frac{1}{2}}$ has no contribution in maximizing the probability of estimated state $x_k$ in \cref{equ:obj_expand}, we incorporate it into the $\gamma$ in \cref{equ:L_k}. Equation \cref{equ:bel_xk_predicted} then becomes 
$
\tilde{x_{k}} = \gamma \exp\{-loss(x_k)\}.
$

Now we look at $loss(x_k)$. Since we already know the function of $loss(x_{k-1},x_k)$, we can obtain the function of $loss(x_k) = L_k - loss(x_{k-1},x_k)$.
The first derivative of $loss(x_k)$ with respect to $x_{k-1}$ is
\begin{equation} \label{equ:1_derivative_loss_xk}
	\begin{split}
		\frac{\partial loss(x_k)}{\partial x_k} =[R_k^{-1}-R_k^{-1}A(A^TR_k^{-1}A + \Sigma_{x_{k-1}}^{-1})^{-1} A^TR_k^{-1}] (x_k-Bu_k)
		-R_k^{-1}A\\(A^TR_k^{-1}A + \Sigma_{x_{k-1}}^{-1})^{-1}\Sigma_{x_{k-1}}^{-1}\mu_{x_{k-1}}.
	\end{split}
\end{equation}
According to the inversion lemma \cite{higham2002}, $(R+PQP^T)^{-1}=R^{-1}-R^{-1}P(Q^{-1}+P^TR^{-1}P)^{-1}P^TR^{-1} $, so we simplify \cref{equ:1_derivative_loss_xk} to
$
	\frac{\partial loss(x_k)}{\partial x_k} = (R_k + A\Sigma_{x_{k-1}}A^T)^{-1} (x_k-Bu_k)	-	R_k^{-1}A(A^TR_k^{-1}A + \Sigma_{x_{k-1}}^{-1})^{-1}\Sigma_{x_{k-1}}^{-1}\mu_{x_{k-1}}.
$

Setting the first derivation to zero gives us the mean value of $x_k$ in $loss(x_k)$:
\begin{subequations}
	\begin{align}
		(R_k& + A\Sigma_{x_{k-1}}A^T)^{-1} (x_k-Bu_k) =R_k^{-1}A(A^TR_k^{-1}A + \Sigma_{x_{k-1}}^{-1})^{-1}\Sigma_{x_{k-1}}^{-1}\mu_{x_{k-1}}\\
		\begin{split}
			x_k &=Bu_k+ (R_k + A\Sigma_{x_{k-1}}A^T)R_k^{-1}A(A^TR_k^{-1}A + \Sigma_{x_{k-1}}^{-1})^{-1}\Sigma_{x_{k-1}}^{-1}\mu_{x_{k-1}}\\
			&=  Bu_k + A(I+ \Sigma_{x_{k-1}}A^TR_k^{-1}A)  (\Sigma_{x_{k-1}}A^TR_k^{-1}A + I)^{-1}\mu_{x_{k-1}} 
			=Bu_k + A\mu_{x_{k-1}}. \label{equ:prediction_mean}
		\end{split}
	\end{align}
\end{subequations}

The second derivation of $loss(x_k)$ leads us to the inverse covariance of $x_k$ in $loss(x_k)$:
\begin{equation} \label{equ:prediction_variance}
		\frac{\partial^2 loss(x_k)}{\partial x_k^2} =(R_k + A\Sigma_{x_{k-1}}A^T)^{-1}=\overline{\Sigma}_{x_k}^{-1}.
\end{equation}

Finally, the prediction $\tilde{x_{k}}$ in \cref{equ:tilde_x} becomes a scaled Gaussian distribution. 
Such distribution has a mean of $\overline{\mu}_{x_k}=Bu_k + A\mu_{x_{k-1}}$ and a covariance of 
$\overline{\Sigma}_{x_k}=R_k + A\Sigma_{x_{k-1}}A^T$. 

Our goal is the solve \cref{equ:obj_expand}. So far we have solved the prediction phase of \cref{equ:tilde_x} in \cref{equ:obj_expand}. Next, we focus on solving $p(y_k|x_k)$ in \cref{equ:obj_expand} that is the measurement phase. 

\subsection{Measurement Phase}
\label{sec:solving_opt_problem}

From \cref{sec:linear_system}, we know that $p(y_k|x_k)$ follows a Gaussian distribution. Furthermore, we concluded that the $\tilde{x_k}$ in \cref{sec:prediction_phase} is also a scaled Gaussian distribution. Thus we can expand \cref{equ:obj_expand} using the Gaussian distribution as 
\begin{equation}
	\begin{split}
	\hat{x} &= \argmax_{x} p(y_k|x_k)\int p(x_k|x_{k-1},u_{k})p(x_{k-1}|y_{1:k-1},u_{1:k-1}) \,dx_{k-1}\\
	&=\argmax_{x} \eta \exp \{ {-\frac{1}{2} (y_k-Cx_k)^T Q_k^{-1} (y_k-Cx_k)} \} \exp \{{-\frac{1}{2} (x_k-\overline{\mu}_{x_k})^T \overline{\Sigma}_{x_k}^{-1} (x_k-\overline{\mu}_{x_k})} \},
	\end{split} 
\end{equation}
where the $\eta$ incorporates the scalar coefficients of the Gaussian distributions, $\overline{\mu}_{x_k}$ and $\overline{\Sigma}_{x_k}$ are the mean and the covariance in \cref{equ:prediction_mean} and \cref{equ:prediction_variance} from prediction phase. 
Let 
\begin{equation} \label{equ:cost_function_jk}
	J_k=\frac{1}{2} (y_k-Cx_k)^T Q_k^{-1} (y_k-Cx_k) + \frac{1}{2} (x_k-\overline{\mu}_{x_k})^T \overline{\Sigma}_{x_k}^{-1} (x_k-\overline{\mu}_{x_k}),
\end{equation}
then the objective function has the form of
$ \label{equ:incoporated_Jk}
	\hat{x} = \argmax_{x} \eta \exp \{ -J_k \}.
$
Because $\eta \exp \{ -J_k \}$ is the product of a Gaussian distribution with a scalar, $\hat{x} = \argmax_{x} \eta \exp \{ -J_k \}$ is the mean value of the distribution. Our goal boils down to obtain the mean value. In addition, in the prediction phase we also utilized the covariance. Therefore we need to calculate the covariance for next iteration as well. 

By calculating the first and second derivatives of \cref{equ:cost_function_jk} with respect to $x_k$, we acquire the mean $\mu_{x_{k}}$ and covariance $\Sigma_{x_{k}}$:
\begin{subequations}
	\begin{align}
		\begin{split} \label{equ:posterior_covariance}
			\Sigma_{x_{k}}^{-1}&=\frac{\partial^2 J}{\partial x_k^2} =C^TQ_k^{-1}C + \overline{\Sigma}_{x_k}^{-1} 
		\end{split}, \hspace{3em}
		\frac{\partial J}{\partial x_k} = -C^TQ_k^{-1}(y_k-Cx_k)+\overline{\Sigma}_{x_k}^{-1}(x_k-\overline{\mu}_{x_k}) =0\\
		\mu_{x_{k}}&=x_k=\overline{\mu}_{x_k} + \Sigma_{x_k}C^TQ_k^{-1}(y_k-C\overline{\mu}_{x_k}) \label{equ:update_eqution}.
	\end{align}
\end{subequations}
So far we get the most probable estimated state $x_k$ in \cref{equ:update_eqution}. 
Moreover we have the covariance in \cref{equ:posterior_covariance} for next iteration. 

In practice solving for the inversion of \cref{equ:posterior_covariance} might be time-consuming. 
To avoid calculating the inversion we change the way to calculate the covariance. 
According to Inversion Lemma \cite{higham2002}, the expression of \cref{equ:posterior_covariance} is rewritten to
$
		\Sigma_{x_k} = (C^TQ_k^{-1}C + \overline{\Sigma}_{x_k}^{-1})^{-1}
		=\overline{\Sigma}_{x_k} - \overline{\Sigma}_{x_k} C^T (Q_k + C\overline{\Sigma}_{x_k}C^T)^{-1}C\overline{\Sigma}_{x_k}
		=[I-\overline{\Sigma}_{x_k} C^T (Q_k + C\overline{\Sigma}_{x_k}C^T)^{-1}C]\overline{\Sigma}_{x_k}
		=(I-K_kC)\overline{\Sigma}_{x_k},
$
where $K_k=\overline{\Sigma}_{x_k} C^T (Q_k + C\overline{\Sigma}_{x_k}C^T)^{-1}$.
Furthermore, it can be proven that $K_k=\overline{\Sigma}_{x_k} C^T (Q_k + C\overline{\Sigma}_{x_k}C^T)^{-1}=\Sigma_{x_k}C^TQ_k^{-1}$, hence \cref{equ:update_eqution} is 
$
	x_k =\overline{\mu}_{x_k} + K_k(y_k-C\overline{\mu}_{x_k}).
$

\paragraph{Summary.}
The goal of KF is to estimate the current state of a system given historical measurements and inputs. 
By utilizing  Bayes rule and MAP, we obtain the objective function of \cref{equ:obj_expand} in a recursive form.
To solve the maximization problem, we split it into two phases, a prediction phase and a measurement phase. 
In the prediction phase we concluded that $\tilde{x_k}$ is Gaussian distributed. Combining such Gaussian distribution with another Gaussian distribution of $p(y_k|x_k)$ in the measurement phase gives us the result of the solution to \cref{equ:obj_expand}.
The KF steps are summarized in \cref{alg:kf}.

\begin{algorithm}[tbh]
  Given $\mu_{x_{k-1}}$, $\Sigma_{x_{k-1}}$, $u_k$, $y_k$ \;
  $\overline{\mu}_{x_k} = Bu_k + A\mu_{x_{k-1}}$\;
  $\overline{\Sigma}_{x_k} = (R_k + A\Sigma_{x_{k-1}}A^T)$\;
  $ K_k=\overline{\Sigma}_{x_k} C^T (Q_k + C\overline{\Sigma}_{x_k}C^T)^{-1}$\;
  $\mu_{x_{k}} =\overline{\mu}_{x_k} + K_k(y_k-C\overline{\mu}_{x_k})$\;
  $\Sigma_{x_k} = (I-K_kC)\overline{\Sigma}_{x_k} $\;
  return $\mu_{x_{k}}$ and $\Sigma_{x_k}$\;
  \caption{The Kalman filter algorithm.}
  \label{alg:kf}
\end{algorithm}

\section{Anomaly Detection for incubator digital twin} \label{sec:anomaly_detection}

In this section we demonstrate the use of KF for anomaly detection for an incubator system. 
In the incubator system, we have the model predicting the states of the system. Furthermore, we have three sensors to measure the room and air temperature inside the incubator.
In the meanwhile, the KF is used to estimate and the states of the system using the measurement data and the predicted states from the DT. 
Furthermore, if an anomaly occurs in the system, the assumption that the process noise is Gaussian no longer holds true. This makes the predictions from KF no longer align with the sensory data from the physical twin indicate. Thus KF can be used for anomaly detection.  

\begin{figure}[h!] 
	\centering
	\includegraphics[width=\linewidth]{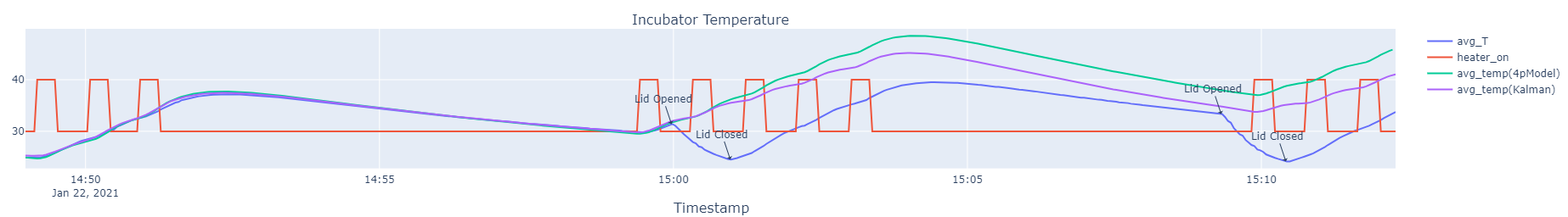}
	\caption{results of anomaly detection for incubator digital twin}
	\label{fig:anomaly_detection}
\end{figure}

In our case, we completely opened the lid of the incubator system when it was running. 
We viewed such an action as an anomaly happening in the system, because the model we used to do the state estimation only represents the behaviors of the system with a closed lid.
Opening it causes the system to violate the physical principles the model was originally built on, on, and will therefore make the KF fail to track what is happening.
The result can be seen in \cref{fig:anomaly_detection}. 
The orange line is the control signal of the heated, \enquote{high} means turning on the heated and \enquote{low} is turning off. 
Green line, purple line, and blue line are predicted state form the model, estimated state from KF, and measurement from the sensors respectively. 

During the experiment, we opened the lid at $14:59:58$ and $15:09:19$. After opening the lid for around $1$ minute, we closed the lid at $15:00:58$ and $15:10:25$ respectively.
As can be seen in \cref{fig:anomaly_detection}, after opening the lid at $14:59:58$, the anomaly was detected since there was a discrepancy between the estimated state from KF and the sensory data. 

After closing the lid, the purple line and the blue line were merging gradually, i.e.\ the discrepancy reduced gradually. Before they converge together we opened the lid again, the discrepancy started to increase again. Should we have waited longer before opening the lid again, the discrepancy would have become virtually zero.

Through the running example, the KF successfully demonstrated the ability of state estimation. Such estimation succeeds in detecting an anomaly in the incubator system. Furthermore, this ability can be extended to other applications such as system monitoring.

\section{Conclusion} \label{sec:conclusion}

In this tutorial, we provided the derivation of the KF from the perspective of probability theory. We started with a linear dynamical system. 
Based on MAP, maximizing the posterior of the state $x_k$ is objective function. We split it into a prediction phase and measurement phase for the derivation of the posteriori. 
We proved that the each of the phases resulted a Gaussian distribution. 
Thus the posterior of the estimated state is Gaussian distributed. 
Such Gaussian distribution enables us to obtain best estimation of state $x_k$ which is the expectation of the distribution. 

We demonstrated the use of KF through a running example, the incubator system. 
We modeled our incubator system as a linear dynamical system in the matrix form. Such form can be substituted into \cref{equ:state_space_equation_noise}. 
We applied the KF to estimate the states of the system and to detect a system anomaly. In the experiments, we completely opened the lid of the incubator system when it was running and the KF successfully detected the anomaly, which is illustrated in \cref{fig:anomaly_detection}.
This feature can be further used for more advance applications.

\section*{Acknowledgments}
We are thankful for the discussions with Yuan Zhao, funded by National Science Foundation of Tianjin 20JCQNJC0039. Hao Feng also acknowledges support from China Scholarship Council.



\bibliographystyle{unsrt}
\bibliography{Mybib,claudio_gen}

\end{document}